\documentstyle[12pt]{article}
\def\b:{\begin{equation}}
\def\e:{\end{equation}}
\def\be:{\begin{eqnarray}}
\def\ee:{\end{eqnarray}}
\def\nn{\nonumber\\}

\def\voo{\vglue 0.5cm}

\def\noi{\noindent}
\def\la{\langle}
\def\ra{\rangle}

\def\sla{\kern -2.2mm /}

\def\bi{\bibitem}

\newcommand{\lb}[1]{\label{eqn:#1}}
\newcommand{\rf}[1]{\ref{eqn:#1}}

\begin{document}

\baselineskip 15pt
\vskip 1.5cm
\begin{flushright}
TMUP-HEL-9507
\end{flushright}
\vskip .5cm
\begin{center}
\begin{Large}
{\bf Phase Space Discretization and Moyal Quantization}
\end{Large}

\vskip 2cm
Ryuji KEMMOKU\footnote[1]{E-mail: kemmoku@phys.metro-u.ac.jp}
 and Satoru SAITO\footnote[2]{E-mail: saito@phys.metro-u.ac.jp}
\vskip  .5cm

{\it Department of Physics, Tokyo Metropolitan University}

{\it Minami-Osawa, Hachioji, Tokyo 192-03, Japan}

\end{center}

\vskip 1.5cm
\centerline{Abstract}
\vskip .5cm
The Moyal quantization is described as a discretization of the classical phase
space by using difference analogue of vector fields.  Difference analogue of
Lie brackets plays the role of Heisenberg commutators.


\vskip 2cm
\noi
{\bf 1. Introduction}
\voo
We have learned in the last few decades the importance of the role played by
the difference version of various integrable nonlinear systems. Some of
examples are: (1) Every soliton type nonlinear differential equation has its
unique difference analogue which preserves integrability \cite{hir}. (2) Many
solvable 2D lattice models have been known to reduce to integrable conformal
field models in the continuous limit \cite{bpz}. (3) The discrete version of
the Calogero-Sutherland-Moser model can be characterized by the Macdonald
symmetric function which is a natural generalization of Jack polynomials
\cite{sogo}.

The existence of difference analogue of integrable systems is not trivial at
all, since in general a naive discretization of continuous variables will not
preserve integrability but creates chaos in an arbitrary nonlinear system. The
transition between integrable and nonintegrable discretizations is subtle and
difficult to clarify the mechanism. It is, however, supposed that a large
symmetry exists behind the integrable discrete system. We know, for instance,
that the quantum group plays an important role in the solvable 2D lattice
models as well as 2D conformal field theories. Hirota's bilinear difference
equation \cite{hir}, which is a difference version of the KP-hierarchy of
soliton equations, exhibits the large symmetry explicitly in the form of the
Pl\"ucker relation. Besides these well known examples we have studied, in a
series of papers \cite{ks}, a discrete version of the Virasoro algebra and
shown that it is nothing but the $W_{1+\infty}$ algebra. We have also
investigated a difference analogue of the logistic equation which consists of
integrable and nonintegrable parts in one model depending on a parameter
\cite{sss}. It was shown that the Julia set characterizing the standard
logistic map disappears when the parameter is adjusted for the model being
integrable and appears conformal symmetry only at this point.

The aim of this article is to show the fact that the discretization of
continuous space discussed above naturally leads us to the Moyal type of
quantization
\cite{moy} when applied to the physical phase space.

The Moyal quantization is one of the ways to describe the quantum mechanics
\cite{moy}. It was proposed based on the work by Wigner who considered
probability function on the phase space in order to study quantum corrections
for thermodynamic equilibrium \cite{wig}. Moyal attempted an interpretation of
the quantum mechanics from the view of general statistical theory. In his
formulation the quantum probability is defined on the physical phase space and
general coordinates and momenta are respected equally. This should be
contrasted with the standard probabilistic interpretation of the quantum
mechanics in which wave functions depend only on half of the variables of the
phase space. To support the new interpretation he introduced the so-called
Moyal bracket which replaces the Poisson bracket in the classical mechanics.
\vskip 1cm
\noi
{\bf 2. Difference Analogue of Vector Fields}
\voo

We would like to show in what follows that the discretization of the phase
space reads to a natural definition of a difference analogue of vector fields.
The Moyal bracket will be shown to follow naturally as a difference version of
the Lie bracket in the place of the Poisson bracket.


The difference operator
\b:
{e^{i\lambda\partial_x}-e^{-i\lambda\partial_x}\over 2i\lambda}={1\over
\lambda}\sin (\lambda\partial_x)
\e:
replaces the derivative
\b:
\partial_x={\partial\over\partial x}
\e:
when the space of the continuous variable $x$ is discretized into a space of
lattice constant $\lambda$. When there are more than one variable we should
have more general expression
\b:
\nabla_{\vec a}:= {1\over \lambda}\sin (\sum_ja_j\partial_{x_j})
\lb{sabun}
\e:
where the lattice vector $\vec a=\{a_j\}$ is proportional to $\lambda$. If the
discretization is not homogeneous the lattice constant differs from one place
to another. This generalization defines the difference analogue of the vector
field
\b:
X^D=\int d\vec a\ v(\vec x,\vec a)\nabla_{\vec a},
\lb{vector field}
\e:
where $v$ is the component of $X^D$ in the local coordinate system $\vec x$
on a manifold which specifies the local dependence of
the lattice constants. Comparing with the ordinary vector fields of the
differential geometry we see that the lattice constant vectors $\vec a$ play
the role of the indices $j$ of the local coordinates $x_j$. The term `space
discretization' corresponds literally to the special case of $v(\vec x,\vec a)$
which is proportional to $\delta (\vec a-\vec a_0)$ with some constant vector
$\vec a_0$. Instead of dealing with such a special case we will be interested
in considering an ensemble of all possible discrete spaces, so that $X^D$ form
a vector space of infinite dimension.

Since a physical quantity is associated with a difference analogue of vector
field which acts on the phase space we must give a prescription to associate a
$c$-number with the operator. This will be done by defining an operator which
is dual of the difference vector field. We are interested in finding a
difference analogue of the differential form and discuss the `difference
geometry'.

To this end we search an operator which invalidates the operation of
$X^D$ on a function.
We now define the `difference one form' $\Omega^D$ by
\b:
\Omega^D := \int d\vec a\ w(\vec x, \vec a)\Delta^{\vec a},
\e:
where $\Delta^{\vec a}$ is the conjugate of $\nabla_{\vec a}$ satisfying the
orthogonality relations under the inner product defined by
\b:
\la \Delta^{\vec a'},\nabla_{\vec a}\ra=\delta(\vec a'-\vec a).
\lb{orthogonal}
\e:
The inner product of $\Omega^D$ with the vector field $X^D$ of $(\rf{vector
field})$ follows as
\be:
\la\Omega^D,X^D\ra&=&\int d\vec a\int d\vec a'\ \left\la  w(\vec x, \vec
a')\Delta^{\vec a'},\ v(\vec x, \vec a)\nabla_{\vec a}\right\ra\nn
&=&\int d\vec a\ w(\vec x, \vec a)v(\vec x, \vec a)
\lb{innerprod}
\ee:
If we consider the following operator
\b:
\lambda\csc(\vec a\cdot\vec\partial) := {2i\lambda\over e^{i\vec
a\cdot\vec\partial}-e^{-i\vec a\cdot\vec\partial}}=\lambda\sum_{n=0}^\infty
e^{-i(2n+1)\vec a\cdot\vec\partial},
\lb{dual}
\e:
it precisely cancells the effect of the operation of $\nabla_{\vec a}$ in
$(\rf{sabun})$.
This enables us to define the bilinear pairing of $(\rf{orthogonal})$
explicitly, for example, by
\be:
\la \Delta^{\vec a'},\nabla_{\vec a}\ra:=\lambda\csc(\vec a'\cdot\vec\partial)
\cdot {1\over\lambda}\sin(\vec a\cdot\vec\partial)\ \delta(\vec a'-\vec a)
=\delta(\vec a'-\vec a).
\ee:
This can be contrasted with the case of the $q$-analysis,
in which the dual of the $q$-difference operator
is defined by the so-called Jackson integral \cite{jac},
$
\int_0^z d_q z'\ f(z'):= z(1-q)\sum_{n=0}^{\infty}f(zq^n)q^n
$,
which can be also expressed in a similar way to $(\rf{dual})$.

The geometrical meaning of the conjugate space spanned by $\Delta^{\vec a}$'s
as a tangent space has been lost. Nevertheless it is a vector space and we can
define `forms' of higher degree so that systems of many particles can be
studied. For example the difference two form is defined naturally by
\b:
\Omega^D=\int d\vec a_1\int d\vec a_2\ w(\vec x, \vec a_1,\vec a_2)\Delta^{\vec
a_1}\wedge \Delta^{\vec a_2}.
\lb{2form}
\e:
Here the wedge product $\wedge$ satisfies the following:
\b:
\Delta^{\vec a_1}\wedge \Delta^{\vec a_2}=-\Delta^{\vec a_2}\wedge \Delta^{\vec
a_1},
\e:
\b:
(\Delta^{\vec a_1}\wedge \Delta^{\vec a_2})
(\nabla_{\vec b_1},\nabla_{\vec b_2})
=\det\la\Delta^{\vec a_i},\nabla_{\vec b_j}\ra\hspace{.3in}(i,j=1,2).
\e:
Moreover we are able to define the `exterior difference operator' $\Delta$.
When it is applied to the difference two form of $(\rf{2form})$, for instance,
it should change the $\Omega^D$ into
\b:
\Delta\Omega^D=\int d\vec a\int d\vec a_1\int d\vec a_2\ \nabla_{\vec a}w(\vec
x,\vec a_1,\vec a_2)\Delta^{\vec a}\wedge\Delta^{\vec a_1}\wedge\Delta^{\vec
a_2}.
\e:
Since $[\nabla_{\vec a_1}, \nabla_{\vec a_2}]=0$, $\Delta$ satisfies the
desired property $\Delta\Delta=0$.

\vskip 1cm
\noi
{\bf 3. Discretization of the Physical Phase Space}
\voo
Now let us consider the physical phase space $\vec x=(p,x)$ whose coordinates
are discretized. We like to establish difference analogue of the Hamilton
vector field
 and the Poisson bracket. In the continuous phase space the Hamilton vector
field $X_f$ operates on a scalar function $g$ as
\b:
X_fg=\biggl({\partial f\over\partial p}\partial_x  -
{\partial f\over\partial x}\partial_p\biggr)\ g.
\lb{Lie}
\e:
We then propose a difference version of the operator $X_f$ :
\b:
X^D_f=\int d\vec a\ v_f(\vec x, \vec a)\nabla_{\vec a},
\lb{dif}
\e:
such that $v(\vec x, \vec a)$ in $(\rf{vector field})$ is given by
\b:
v_f(\vec x,\vec a)={1\over (2\pi \lambda)^2}\int d\vec b\
e^{i{1\over\lambda}(\vec a\times\vec b)}f(\vec x +i\vec b).
\e:
In this expression ${1\over \lambda}(\vec a\times\vec b)$ is the area in the
unit of $\lambda$ of the parallelogram formed in the phase space by the two
vectors $\vec a$ and $\vec b$. The symplectic structure in $(\rf{Lie})$ is
retained in this expression as the symmetry under the exchange of these vectors
$\vec a$ and $\vec b$.

If we perform the integrations we have the following simple expression for
$X^D_fg$ :
\b:
{1\over \lambda}\sin\{\lambda(\vec\partial_1\times\vec\partial_2)\}
f(\vec x_1)g(\vec x_2)\nn
:= \{f,g\}_M
\lb{Moyal}
\e:
where $\vec x_j=\vec x,\ \ j=1,2$ are implied after all calculations.
$(\rf{Moyal})$ is nothing but the Moyal bracket for the functions $f$ and $g$
defined on the phase space. In other words our $X^D_fg$ can be regarded as a
difference operator representation of the Moyal bracket. As is well known the
Moyal bracket reduces into the Poisson bracket in the classical limit $\hbar
\rightarrow 0$. In our formulation, it is realized by taking the $\lambda \to
0$ limit in $(\rf{Moyal})$ and the phase space becomes continuous.

Since we have the difference analogue of the Hamilton vector field we ask if
there exists a difference analogue of the Lie bracket which closes an algebra.
This is a highly nontrivial question since the difference operator
$(\rf{sabun})$ is not a local operator. Therefore it is remarkable that $X^D_f$
satisfies the following commutation relation:
\b:
[X^D_f, X^D_g]=X^D_{\{f,g\}_M}.
\lb{comm}
\e:
Hence $X^D_f$ is a generator of an infinite dimensional Lie algebra.

Some comments are in order:

\noi
(1) In the series of papers \cite{ffz} Fairlie {\it et al.}\  studied the same
algebra as $(\rf{comm})$ generated by the shift operators
\be:
K_f&=&{1\over 2i\lambda}f(p-i\lambda\partial_x,x+i\lambda\partial_p)\nn
&=&{1\over 2i\lambda}\sum_{m,n\in {\bf Z}}f_{m,n}\exp
(imp+inx)\exp\left(\lambda m\partial_x-\lambda n\partial_p\right)
\ee:
where $f(p,x)=\sum_{m,n}f_{m,n}\exp (imp+inx)$. Our algebra is a subalgebra of
theirs which consists of only antisymmetric combinations of the generators. We
emphasize that these particular combinations, which are direct difference
analogue of the Hamilton vector field, close the algebra among themselves.

\noi
(2) Our algebra $(\rf{comm})$ is an algebra which is satisfied by the Moyal
bracket. We gave a representation of the generators which enables us to
interpret them as difference analogue of the Hamilton vector field. There have
been some geometrical arguments of the Moyal bracket \cite{bay}, but all based
on the ordinary differential geometry. Our view of the Moyal bracket stands on
a completely different geometry.

\noi
(3) We have pointed out in our previous work \cite{ks} the equivalence of the
$W_{1+\infty}$ algebra and the discrete version of the Virasoro algebra. This
correspondence is established again in the above algebra within more general
context.

\noi
(4) Since $X^D$ form a Lie algebra of infinite dimension it is likely that
there exists an integrable system associated with it.
\vskip 1cm
\noi
{\bf 4. Quantum Mechanics}
\voo
The ordinary procedure of quantization is achieved by replacing a Poisson
bracket defined in the classical phase space by a commutator of operators which
act on a Hilbert space. Moyal has shown that this procedure is equivalent to
replacing the Poisson bracket by a Moyal bracket. This correspondence of the
classical and the quantum mechanics is somehow mysterious since there exists no
explanation how such a transition takes place in the nature. On the other hand
we have started from defining the difference analogue of derivatives and
derived the difference analogue of the Lie bracket which forms the algebra
generated by the Moyal brackets. Therefore we can interpret the gap between the
classical and quantum mechanics such that the quantization is realized by a
discretization of the classical phase space.

In order to accomplish the correspondence let us identify the lattice constant
$\lambda$ with the Planck constant $\hbar$. Then the Poisson bracket of the
physical quantities defined in the classical phase space have to be replaced by
the commutator $(\rf{comm})$ since $\hbar$ is finite. This is our new procedure
of quantization.

According to this prescription of the quantization, the time evolution of a
physical variable $A(\vec x)$ should follow to
\b:
i\hbar{d\over dt}X^D_A=[X^D_A, X^D_H],
\lb{equation of motion}
\e:
when the system is governed by a static Hamiltonian $H$. Because of
$(\rf{comm})$ this is equivalent to the following statement in the language of
Moyal bracket;
\b:
i\hbar{d\over dt}A=\{A, H\}_M.
\lb{moyeq}
\e:

Now we like to establish the way to associate our operators with physical
observation. In the statistical theory an average of observation of a dynamical
variable $A$ defined as a function in the phase space $\vec x=(p,x)$ is given
by $\int d\vec x\ F(\vec x)A(\vec x)$, where $F(\vec x)$ is the distribution
function of the system under consideration which satisfies $\int d\vec x \
F(\vec x)=1$. We have already provided a way to associate an operator $X^D$ to
a $c$ number. The problem to be solved here is to find an operator which
represents the distribution function $F(\vec x)$. We will solve it as follows.
First we notice that if we choose $w(\vec x,\vec a)$ of $\Omega^D$ in
$(\rf{innerprod})$ as
\b:
w(\vec x,\vec a)={1\over(2\pi\lambda)^2}\int d\vec b\ e^{-{i\over\lambda}\vec
a\times\vec b}f^*(\vec x+i\vec b),
\e:
with $^*$ the complex conjugation, the inner product with $X^D_g$ becomes $\int
d\vec a\ \tilde f^*(\vec a)\tilde g(\vec a)$. Here $\tilde f(\vec a)$ is the
Fourier transform of $f(\vec x)$. Therefore if we define the one form $P_F$ by
\b:
P_F=\int d\vec a\int d\vec b\ e^{-{i\over\hbar}\vec a\times\vec b}F^*(\vec
x+i\vec b)\Delta^{\vec a},
\e:
we obtain the expectation value of $A(\vec x)$ as
\be:
\la P_F, X^D_A\ra&=&(2\pi\hbar)^2\int d\vec a\ \tilde F(\vec a)\tilde A(\vec a)
\nn
&=&\int d\vec x\ F(\vec x)A(\vec x).
\lb{expect}
\ee:
Here we used the fact that the distribution function $F(\vec x)$ is real.
{}From this argument we conclude that $\la P_F, X^D_A\ra$ is the rule to
evaluate the expectation value of physical quantities in our framework.

Now the correspondence between the ordinary quantum mechanics and our formalism
must be clarified. For this purpose we recall that in the ordinary quantum
mechanics the expectation value of an observable ${\bf A}$ in the pure state
$|\psi\ra$ is expressed by $\la\psi|{\bf A}|\psi\ra$. According to the Weyl
correspondence \cite{moy}, we can write
\be:
{\bf A}(\vec x)&=&\int d\vec a\ \tilde A(\vec a)e^{{i\over\hbar}\vec a \cdot
\vec {\bf x}}\\
\tilde F(\vec a)&=&\la \psi|e^{{i\over\hbar}\vec a\cdot\vec{\bf x}}|\psi\ra.
\ee:
Substituting these into $(2\pi\hbar)^2\int d\vec a\tilde F(\vec a)\tilde A(\vec
a)$ of $(\rf{expect})$ we obtain
\b:
\la P_F, X^D_A\ra=\la\psi|{\bf A}|\psi\ra.
\e:

By use of the definition, we can consider the time evolution of the state
density in the phase space.  The time dependence of $\la P_F,X_A^D\ra$ can be
two fold: in the Heisenberg picture we have $\la P_F,X_A^D\ra_t=\la
P_F,X_A^D(t)\ra$, while in the Schr\"odinger picture it is given by $\la
P_F(t),X_A^D\ra$, which must be equivalent.
(In the following discussion, we assume that $P_F$ (in Heisenberg) and
$X_A^D$ (in Schr\"odinger) have no explicit time dependence.)
Then the solution of $(\rf{equation of motion})$ becomes
\b:
X_A^D(t)=e^{iX_H^Dt/\hbar}X_A^D e^{-iX_H^Dt/\hbar}
\e:
This implies that the solution of $(\rf{moyeq})$ can be expressed as
\b:
A(t)=e^{iX_H^Dt/\hbar}A.
\e:
In the Heisenberg picture,
\b:
i\hbar{d\over dt}\la P_F,X_A^D(t)\ra=\la P_F,[X^D_A(t),X_H^D]\ra
=\la P_F,X^D_{\{A(t),H\}_M}\ra.
\e:
We can show that the right hand side is identical with
$\la P_{\{H,F(t)\}_M},X^D_A\ra$
if we define
\b:
F(t):=e^{-iX_H^Dt/\hbar}F.
\e:
 Hence we obtain the relation which must be hold by the state density operator
$P_F(t)$ ;
\b:
i\hbar{d\over dt}P_F(t)=P_{\{H,F(t)\}_M}.
\e:
This determines the time evolution of the quantum state in the Schr\"odinger
picture.
The above equation is equivalent to
\b:
i\hbar{d\over dt}F(t)=\{H, F(t)\}_M,
\e:
which is nothing but the equation of motion for the state density in the Moyal
formalism \cite{moy,wig}.

We have shown that the quantum mechanics can be reformulated in terms of
difference analogue of the Hamilton vector field. This formalism is described
mostly in parallel with one of Moyal and reproduces the same results. The merit
of adding new formalism is that it is based on the discretization of the
physical phase space and fills the gap between the classical and quantum
mechanics. The Planck constant is introduced as the unit of phase space
discretization. From more practical point of view we are interested in the
basic algebra $(\rf{comm})$ which holds among operators associated with quantum
mechanical observables. We know that it is satisfied by the generators of the
discrete analogue of the Virasoro algebra \cite{ks}. Therefore we expect that
the algebra will characterize certain class of integrable systems. Results
along this direction of study should be reported in our forthcoming papers.

This work is supported by Fiscal Year
1994 Fund for Special Research Project at Tokyo Metropolitan University and the
Grant-in-Aid for general Scientific Research from the Ministry of
Education, Science and Culture, Japan (No.02640234)


\end{document}